\documentclass[conf]{new-aiaa} 
\usepackage[utf8]{inputenc}

\usepackage{graphicx}

\usepackage{amsmath}
\usepackage{amsmath} 
\usepackage{amsfonts}
\newtheorem{proposition}{Proposition}[section]

\usepackage{float} 

\usepackage{subcaption}

\usepackage{tikz}
\usetikzlibrary{circuits.ee.IEC}
\usetikzlibrary{shapes,arrows,calc,positioning}
\tikzstyle{bigblock} = [draw, fill=blue!20, rectangle, 
    minimum height=6em, minimum width=8em]
\tikzstyle{medblock} = [draw, fill=blue!20, rectangle, 
    minimum height=4em, minimum width=4em]    
\tikzstyle{mux} = [draw, fill=black!20, rectangle, 
    minimum height=5em, minimum width=0.1em]    
\tikzstyle{smallblock} = [draw, fill=blue!20, rectangle, 
    minimum height=2em, minimum width=3em]
    
\tikzstyle{data_block} = [draw, fill=green!20, rectangle, 
    minimum height=2em, minimum width=3em]
\tikzstyle{ops_block} = [draw, fill=blue!20, rectangle, 
    minimum height=2em, minimum width=3em]    
\tikzstyle{est_block} = [draw, fill=red!20, rectangle, 
    minimum height=2em, minimum width=3em]    
    
\tikzstyle{sum} = [draw, fill=blue!20, circle, node distance=1cm,minimum height=0.5cm]
\tikzstyle{signal} = [coordinate]
\tikzstyle{pinstyle} = [pin edge={to-,thin,black}]
\tikzstyle{block} = [draw, fill=blue!20, rectangle, 
    minimum height=3em, minimum width=9em]
\tikzstyle{blockS} = [draw, fill=blue!20, rectangle, 
    minimum height=3em, minimum width=4em]    
\tikzstyle{input} = [coordinate]
\tikzstyle{output} = [coordinate]

\tikzstyle{gain} = [draw, fill=blue!20, regular polygon, regular polygon sides=3, shape border rotate=30]
\tikzstyle{gain_vert} = [draw, fill=blue!20, regular polygon, regular polygon sides=3, shape border rotate=0]

\usetikzlibrary{matrix}

\usetikzlibrary{positioning}
\usetikzlibrary{math}

\usepackage{hyperref}
\usepackage{xcolor}
\hypersetup{
    colorlinks,
    linkcolor={blue!100!black},
    citecolor={blue!50!black},
    urlcolor={blue!80!black}
}

\newcommand{\bc}{\begin{center}}
\newcommand{\ec}{\end{center}}
\newcommand{\benum}{\begin{enumerate}}
\newcommand{\eenum}{\end{enumerate}}

\newcommand{\matl}{\left[ \begin{array}}
\newcommand{\matr}{\end{array} \right]}

\renewcommand{\matl}{\begin{bmatrix}}
\renewcommand{\matr}{\end{bmatrix}}

\newcommand{\matls}{\left[ \begin{smallmatrix}}
\newcommand{\matrs}{\end{smallmatrix} \right]}
\newcommand{\isdef}{\stackrel{\triangle}{=}}

\newcommand{\rmT}{{\rm T}}

\newcommand{\rmb}{{\rm b}}
\newcommand{\rmc}{{\rm c}}

\newcommand{\rme}{{\rm e}}
\newcommand{\rmf}{{\rm f}}

\newcommand{\rmr}{{\rm r}}
\newcommand{\rms}{{\rm s}}

\newcommand{\BBC}{{\mathbb C}}

\newcommand{\BBR}{{\mathbb R}}

\newcommand{\SM}{{\mathcal M}}
\newcommand{\SN}{{\mathcal N}}

\newcommand{\shiftq}{{\textbf{\textrm{q}}}}

\usepackage{enumitem,amssymb}
\newlist{todolist}{itemize}{2}
\setlist[todolist]{label=$\square$}
\usepackage{pifont}

\usepackage[colorinlistoftodos,textwidth=1 in,textsize=footnotesize]{todonotes}
\usepackage{setspace}

\usepackage{xcolor}
\usepackage[normalem]{ulem}
\usepackage{bm}
\usepackage{setspace}

\title{Data-driven Pressure Recovery in Diffusers}

\author{
Juan Augusto Paredes Salazar\footnote{Postdoctoral Research Fellow,  \texttt{japarede@umbc.edu}}, 
Ankit Goel\footnote{Assistant Professor, \texttt{ankgoel@umbc.edu}}
}
\affil{Department of Mechanical Engineering, \\
University of Maryland Baltimore County, Baltimore, MD 21250.}
\author{
  Rowen Costich\footnote{Undergraduate Researcher in the Hypersonic Aerothermal Vehicle Analysis (HAVA) Laboratory, \texttt{costir@rpi.edu} }, 
  Meliksah Koca\footnote{Graduate Research Assistant in the Hypersonic Aerothermal Vehicle Analysis Laboratory \texttt{kocam@rpi.edu}}, 
  Ozgur Tumuklu\footnote{Assistant Professor, AIAA Senior Member, PI of the Hypersonic Aerothermal Vehicle Analysis Laboratory, \texttt{tumuko@rpi.edu} }, 
  Michael Amitay\footnote{Professor and James L. Decker 45 Endowed Chair in Aerospace Engineering, and Director of the Center for Flow Physics and Control} 
  }
  \affil{Mechanical, Aerospace and Nuclear Engineering, \\
  Rensselaer Polytechnic Institute, Troy, NY 12180, USA}

\begin{document}
 \doublespacing
\maketitle
\begin{abstract}
This paper investigates the application of a data-driven technique based on retrospective cost optimization to optimize the frequency of mass injection into an S-shaped diffuser, with the objective of maximizing the pressure recovery.
Experimental data indicated that there is an optimal injection frequency between 100 Hz and 300 Hz with a mass flow rate of 1 percent of the free stream. 
High-fidelity numerical simulations using compressible unsteady Reynolds-Averaged Navier-Stokes (URANS) are conducted to investigate the mean and temporal features resulting from mass injection into an S-shaped diffuser with differing injection speeds and pulse frequencies.
The results are compared with experiments to confirm the accuracy of the numerical solution.
Overall, 2-D simulations are relatively in good agreement with the experiment, with 3-D simulations currently under investigation to benchmark the effect of spanwise instabilities. 
Simulation results with the proposed data-driven technique show improvements upon a baseline case by increasing pressure recovery and reducing the region of flow recirculation within the diffuser.
%
\end{abstract}

\section{Introduction}

The design of subsonic and supersonic diffusers is critical in various engineering applications, including engine intakes, hydraulic systems, turbines, wind tunnels, and HVAC systems \cite{zhang2000automated, harloff1996supersonic, lee1985subsonic, mehta1979aerodynamic}.
The critical function of a diffuser is to decelerate the flow passing through it while minimizing the pressure losses. 
%
Subsonic and supersonic diffusers differ fundamentally in their designs.
Whereas a subsonic diffuser expands in the flow direction to decelerate the flow, the supersonic and hypersonic diffusers reduce the area to decelerate the flow.
%

In supersonic diffusers, decelerating the flow in the presence of inevitable shock waves presents greater challenges, as shock waves originating from other surfaces or reflected from boundaries can impinge on the inlet surface, potentially causing flow separation \cite{babinsky2011shock}. 
Short, curved inlet ducts have been of immense interest, as they allow a reduction in the aircraft's overall length and enable embedding the engine within the airframe's body, resulting in a significant weight reduction.
However, the rapid inlet curve results in a large area of flow separation and secondary flow structures within the duct, reducing pressure recovery and causing unstable conditions at the aerodynamic interface plane. In fact, the presence of massive separation regions and secondary structures reduces the effectiveness of diffusers and introduces losses.

Over the last few decades, several flow-control strategies have been investigated to mitigate pressure losses and improve diffuser efficiency \cite{o2020active, gartner2019mitigation}. 
Such flow-control approaches can be classified as passive or active strategies. 
In passive flow-control strategies, the geometry of the physical diffuser is carefully modified to achieve a desired performance. 
For example, vortex generators are widely used passive flow-control devices that delay the flow separation over wings. 
However, due to their passive nature, such strategies are only suitable/beneficial in a narrow operational range. 
In contrast, active flow-control strategies rely on feedback control, where a dynamic system uses real-time measurements to adjust control inputs and achieve the desired flow behavior.
Designing such feedback control systems is particularly challenging, as the underlying fluid-dynamics models are often poorly suited to control-oriented formulations.
Furthermore, the flow physics is high-dimensional, nonlinear, and highly transient, which makes the control system design extremely challenging. 
State-of-the-art model-based active flow-control strategies typically rely on reduced-order models.
However, these models are only accurate within a limited operating regime, which often limits the effectiveness and robustness of the resulting control systems.
Given these challenges, data-driven and learning-based adaptive control approaches have emerged as promising alternatives for addressing the complexities of flow control.

In this work, we investigate the application of retrospective cost adaptive control (RCAC) to maximize the pressure recovery in a supersonic diffuser. 
RCAC is a discrete-time, data-driven, adaptive control technique for stabilization, command-following, and disturbance-rejection problems. 
It requires minimal modeling information and uses an online recursive optimization algorithm to adjust the controller coefficients based on continuously measured data. 
An overview of RCAC is provided in \cite{rahman2017retrospective} and its applications to a wide variety of engineering problems such as helicopter vibration reduction, battery health monitoring, spacecraft attitude control, scramjet thrust regulation, noise suppression, and electricity grid regulation are described in \cite{padthe2013retrospective, zhou2013battery, camblor2014retrospective, goel2015scramjet, goel2018retrospective,goel2019output, xie2019adaptive, islam2019phasor}.
RCAC offers several features that make it particularly well-suited for flow-control problems. 
It can adapt in real time and optimize arbitrary controller architectures online without requiring a detailed system model \cite{goel2021experimental}. 
Because RCAC relies solely on measured data, it can be directly integrated with numerical simulations for controller tuning and stress testing.
These characteristics allow RCAC to be initially tuned using a simplified or nominal numerical model and then seamlessly adapt to more complex simulations or the actual physical system.
%
%
%
%
Consequently, in this work, we use the RCAC framework to recursively optimize a control system to minimize the pressure losses in a diffuser.

The paper is organized as follows.
Section \ref{sec:model} describes the governing equations and the computational model of the diffuser considered in this work and presents preliminary results indicating the efficacy of a controllable jet in minimizing pressure loss, 
Section \ref{sec:RCAC_Algo} reviews the RCAC algorithm and discusses implementation details for interfacing RCAC with the diffuser model simulation, and 
Section \ref{sec:results} shows the numerical results from implementing RCAC within the diffuser model simulation.
%
%
Finally, the paper concludes by discussing the results and future work in Section \ref{sec:conclusions}.

\section{Numerical Model and Diffuser Configuration}
\label{sec:model}
This section presents the governing equations of the flow, the numerical framework, and the diffuser geometry considered in this work. 

\subsection{Governing Equations}

The flow inside the diffuser is modeled using the unsteady, two-dimensional, compressible Reynolds-averaged Navier--Stokes (RANS) equations. All simulations are performed with the \texttt{rhoPimpleFoam} solver of OpenFOAM under the RAS turbulence modeling framework~\cite{openfoam,openfoam_foundation}. The turbulence closure is provided by the Menter $k$--$\omega$ SST model~\cite{menter1994two,menter2003sst}, using its standard OpenFOAM implementation~\cite{openfoam}.

\subsubsection{Continuity and Momentum}

The continuity equation is given by
\begin{equation}
\frac{\partial \rho}{\partial t} + \nabla \cdot (\rho \mathbf{U}) = 0,
\end{equation}
where $\rho$ is the fluid density and $\mathbf{U}$ is the velocity vector.
The momentum equation is written as
\begin{equation}
\frac{\partial (\rho \mathbf{U})}{\partial t}
+ \nabla \cdot (\rho \mathbf{U} \otimes \mathbf{U})
= -\nabla p + \nabla \cdot \bm{\tau},
\end{equation}
where $p$ is the static pressure and the viscous stress tensor $\bm{\tau}$ is defined as
\begin{equation}
\bm{\tau}
=
\mu_{\text{eff}}
\left[
\nabla \mathbf{U} + (\nabla \mathbf{U})^{\mathrm{T}}
\right]
-
\frac{2}{3}\mu_{\text{eff}}(\nabla \cdot \mathbf{U})\mathbf{I}.
\end{equation}
Here, $\mathbf{I}$ is the identity tensor and
\begin{equation}
\mu_{\text{eff}} = \mu + \mu_t
\end{equation}
is the effective dynamic viscosity, composed of the laminar dynamic viscosity $\mu$ and the turbulent dynamic viscosity $\mu_t$ obtained from the $k$--$\omega$ SST model.

\subsubsection{Energy Equation}

The total energy equation solved by \texttt{rhoPimpleFoam} is written in conservative form as
\begin{equation}
\frac{\partial (\rho E)}{\partial t}
+ \nabla \cdot \left[ \mathbf{U} (\rho E + p) \right]
= \nabla \cdot (\bm{\tau} \cdot \mathbf{U})
- \nabla \cdot \mathbf{q},
\end{equation}
where the total energy per unit mass is
\begin{equation}
E = C_v T + \frac{1}{2}|\mathbf{U}|^2.
\end{equation}
Here, $T$ is the temperature and $C_v$ is the specific heat at constant volume; the second term represents the kinetic energy per unit mass. The heat flux vector $\mathbf{q}$ is modeled as
\begin{equation}
\mathbf{q} = -k_{\text{eff}} \nabla T, \qquad
k_{\text{eff}} = k + k_t.
\end{equation}
In this expression, $k_{\text{eff}}$ is the effective thermal conductivity, $k$ is the laminar thermal conductivity, and $k_t$ is the turbulent thermal conductivity, which is related to the turbulent viscosity by
\begin{equation}
  k_t = \frac{\mu_t C_p}{\Pr_t},
\end{equation}
where $C_p$ is the specific heat at constant pressure and $\Pr_t$ is the turbulent Prandtl number.

\subsubsection{Equation of State}

The working fluid is modeled as a calorically perfect ideal gas:
\begin{equation}
p = \rho R T,
\end{equation}
where $R$ is the specific gas constant and the specific heats at constant pressure and volume, $C_p$ and $C_v$, are assumed constant, with $R = C_p - C_v$.

\subsubsection{Turbulence Model: $k$--$\omega$ SST}

Turbulence effects are modeled using the Menter $k$--$\omega$ SST model within the RAS framework~\cite{menter1994two,menter2003sst,openfoam}. The transport equations for the turbulence kinetic energy $k$ and the specific dissipation rate $\omega$ are given by
\begin{equation}
\frac{\partial (\rho k)}{\partial t}
+ \nabla \cdot (\rho \mathbf{U} k)
= \nabla \cdot \left[ (\mu + \sigma_k \mu_t)\nabla k \right]
+ P_k - \beta^* \rho k \omega,
\end{equation}
\begin{equation}
\frac{\partial (\rho \omega)}{\partial t}
+ \nabla \cdot (\rho \mathbf{U} \omega)
= \nabla \cdot \left[ (\mu + \sigma_\omega \mu_t)\nabla \omega \right]
+ \alpha \frac{\omega}{k} P_k
- \beta \rho \omega^2 + D_\omega.
\end{equation}
Here, $k$ is the turbulent kinetic energy, $\omega$ is the specific dissipation rate, and $P_k$ denotes the production term of turbulent kinetic energy. The term $D_\omega$ represents the cross-diffusion contribution specific to the SST formulation. The coefficients $\sigma_k$, $\sigma_\omega$, $\alpha$, $\beta$, and $\beta^*$ are standard model constants of the $k$--$\omega$ SST model~\cite{menter1994two,menter2003sst}.

In the OpenFOAM implementation, the turbulent kinematic viscosity $\nu_t$ is computed as~\cite{openfoam}
\begin{equation}
\nu_t = \frac{a_1 k}{\max \big(a_1 \omega,\; b_1 F_2 S \big)},
\end{equation}
where $a_1$ and $b_1$ are model constants, $F_2$ is the SST blending function, and $S$ is the magnitude of the mean strain-rate tensor. The strain-rate tensor is defined as
\begin{equation}
S_{ij} = \frac{1}{2}\left(
\frac{\partial U_i}{\partial x_j} + \frac{\partial U_j}{\partial x_i}
\right),
\end{equation}
and its magnitude is
\begin{equation}
S = \sqrt{2 S_{ij} S_{ij}}.
\end{equation}
The turbulent dynamic viscosity then follows as
\begin{equation}
\mu_t = \rho \nu_t.
\end{equation}

\subsubsection{Wall Modelling}

Wall-bounded turbulence is treated using the standard RAS wall-modelling approach in OpenFOAM~\cite{openfoam}. For high-Reynolds-number simulations, the near-wall region is modeled via wall functions:
\begin{itemize}
  \item the turbulent viscosity at walls is prescribed using a momentum wall function (e.g., \texttt{nutkWallFunction}),
  \item the turbulence variables $k$ and $\omega$ use \texttt{kqRWallFunction} and \texttt{omegaWallFunction}, respectively,
\end{itemize}
which are consistent with the $k$--$\omega$ SST model and the assumed $y^+$ range. These wall functions impose the appropriate logarithmic velocity profile and turbulence behavior close to the solid walls and are recommended for high-Reynolds-number RANS simulations in complex internal flows~\cite{openfoam}.

\subsection{Numerical Solver}

All simulations are performed using the transient, compressible RANS solver \texttt{rhoPimpleFoam}~\cite{openfoam,openfoam_foundation}. The solver employs the PIMPLE algorithm, which combines the PISO and SIMPLE procedures to handle strong pressure--velocity coupling in unsteady flows. Density and temperature are updated consistently with the ideal-gas equation of state, and the effective transport coefficients $\mu_\text{eff}$ and $k_\text{eff}$ are provided by the RAS turbulence model described above.

Spatial discretization uses a finite-volume method on an unstructured mesh. Gradients are approximated with a second-order Gauss linear scheme, convective terms are discretized using a limited linear-upwind scheme, and diffusive terms use a Gauss linear corrected scheme. Time integration is performed with a first-order implicit Euler scheme, subject to CFL constraints associated with the local flow speed and mesh resolution. The turbulence model is selected through the RAS configuration in the \texttt{turbulenceProperties} dictionary, where the $k$--$\omega$ SST model and its associated wall functions are activated according to the OpenFOAM RAS turbulence modelling guidelines~\cite{openfoam,openfoam_foundation}.

\subsection{Diffuser Geometry and Control Parameters}

\begin{figure}[h!]
    \centering
    \begin{tabular}{c}
    \includegraphics[clip=true,trim=04 04 04 04, width=0.85\textwidth]{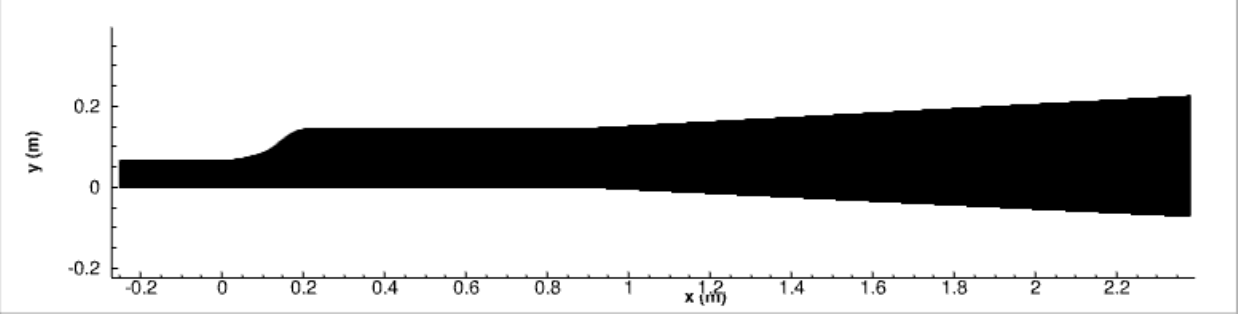}\\
     (a) Computational domain \\
    \includegraphics[clip=true,trim=04 04 04 04, width=0.5\textwidth]{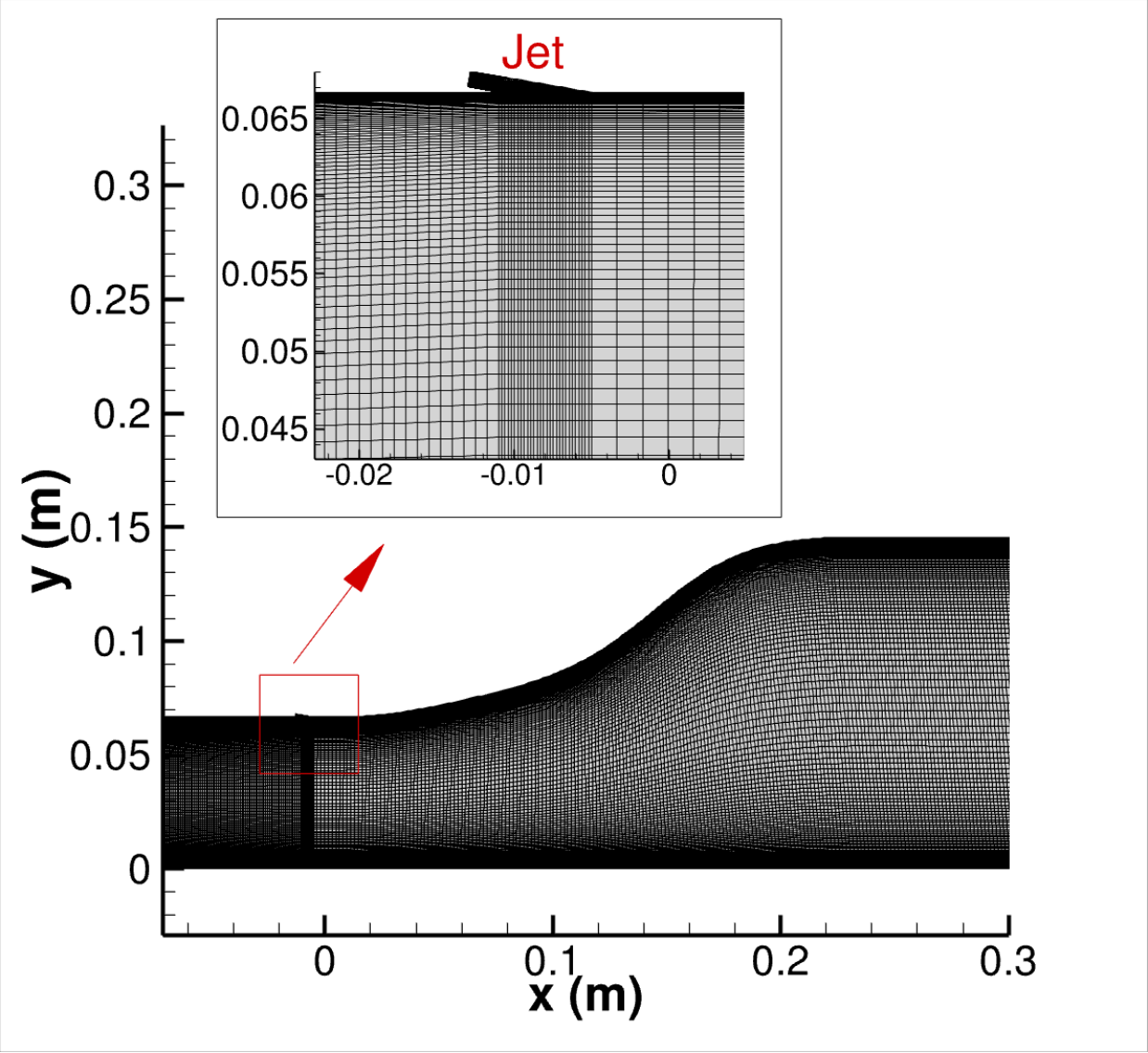}
     (b) 
    \end{tabular}
    \caption{Computational domain for the diffuser along with the location of the control jet and grid points.} 
    \label{fig:ControlDomain}
\end{figure}

\begin{figure}[h]
    \centering
    \includegraphics[clip=true,trim=04 04 04 04, width=0.75\textwidth]{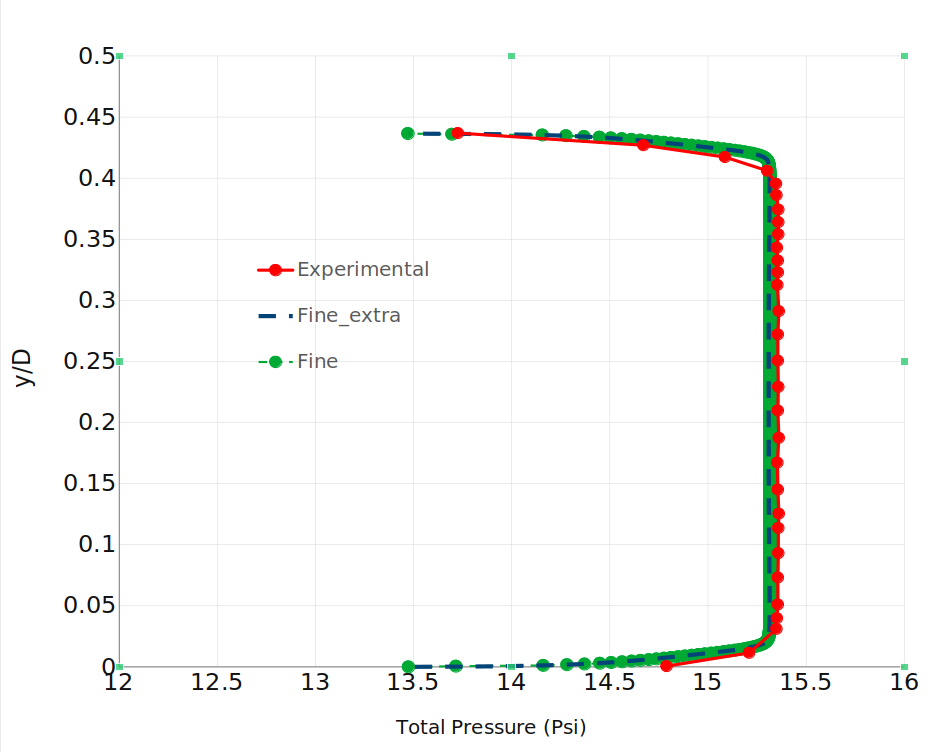}
    \caption{Comparison of the pressure values at the inlet using two grid resolutions with the experiment~\cite{gartner2019mitigation}. } 
    \label{fig:NoControlInletPressure}
\end{figure}

The geometry of the subsonic diffuser, the simulation domain, and the grid points considered in this paper are shown in Figure \ref{fig:ControlDomain}. As shown, the grid resolution, especially near the nozzle lip and walls, is increased to capture the boundary layer. Gmsh~\cite{gmsh,gmsh2} software is used to create the computational mesh. A thorough grid convergence study was performed for the baseline case, and the total pressure at the inlet was compared with the measurements. As shown in Fig. \ref{fig:NoControlInletPressure}, our findings with two different grid levels compare well with the measurements, indicating that our results are independent of grid resolution. 

Figure~\ref{fig:ControlDomain} shows the location of a controllable jet at $x = -0.01$, with a user-specified velocity. The time-varying velocity of the jet is defined as:
\begin{equation}
    \mathbf{V}(t) = \mathbf{L} + A \cdot \sin(2\pi f t) \cdot \mathbf{S}
    \label{Eq:Jetparam}
\end{equation}
where $\mathbf{V}(t)$ is the velocity vector at time \( t \), 
$\bm{L}$ is a constant vector offset (level), 
$A$ is  the amplitude of the sine wave, 
$f$ is the frequency in Hz, 
$t$ is  time in seconds, 
$\bm{S}$ is a scale (direction) vector.
Table~\ref{tab:jetParameters} lists the jet parameters used in this work.

\begin{table}[h!]
\centering
\caption{ Unsteady jet parameters defined in Eq.\eqref{Eq:Jetparam}~\cite{gartner2019mitigation}.}
\begin{tabular}{|c|c|c|c|c|}
\hline
Case & $f$  (Hz) & $\mathbf{L}$ (m/s)&  $\mathbf{S}$ (m/s)  &  $A$ (m/s) \\
\hline
High-Amplitude &200   & (122.99,  -22.35,  0) & (0.99,  -0.179,  0)&110 \\
Low-Amplitude &200  & (71.33,  -12.96,  0)  & (0.99,  -0.179,  0) &57.5 \\
\hline
\end{tabular}

\label{tab:jetParameters}
\end{table}

\subsection{Results and Discussion}
We performed 2-D simulations with steady and pulsed jets located near $x=0$ on the top surface of the diffuser in order to investigate the spatial and temporal characteristics of the flow field. It should be emphasized that the flow is inherently 3-D, as the height and width of the diffuser are comparable. Furthermore, the presence of a large separation region makes the flow susceptible to 3-D secondary structures~\cite{gartner2019mitigation}. 
Full 3-D simulations are ongoing, and our preliminary results reveal the formation of secondary structures near the separation region, which will be the focus of suppression efforts using the  \textcolor{black}{the data-driven, learning-based adaptive control}. 
Nonetheless, the 2-D model captures a significant portion of the flow instabilities while offering a more computationally efficient approach for the early design of the controller. 

\begin{figure}
\centering
\begin{tabular}{cc}
\includegraphics[clip=true,trim=02 02 02 02, width=0.5\textwidth]{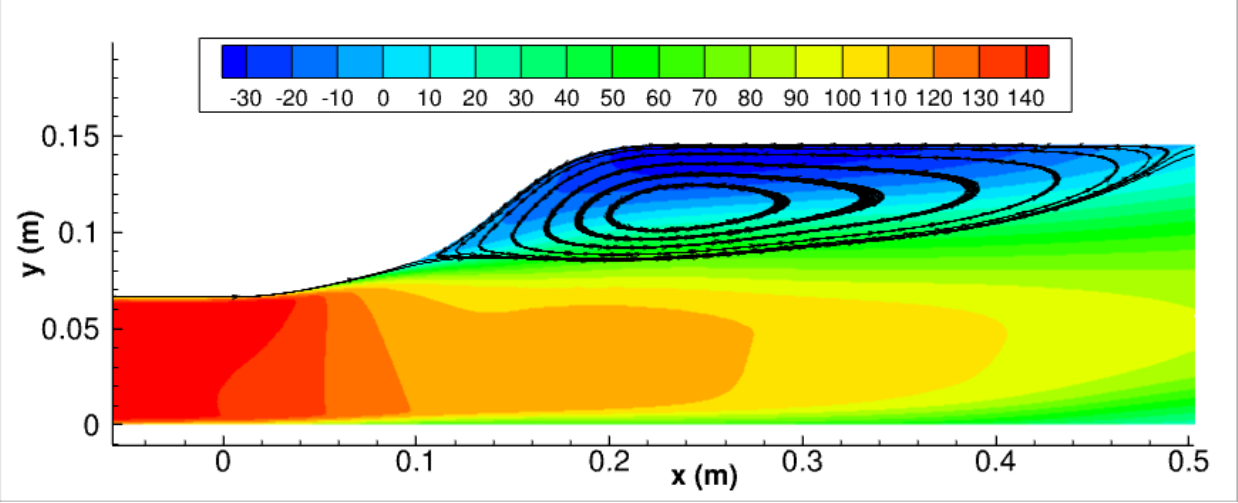}&
\includegraphics[clip=true,trim=02 02 02 02, width=0.5\textwidth]{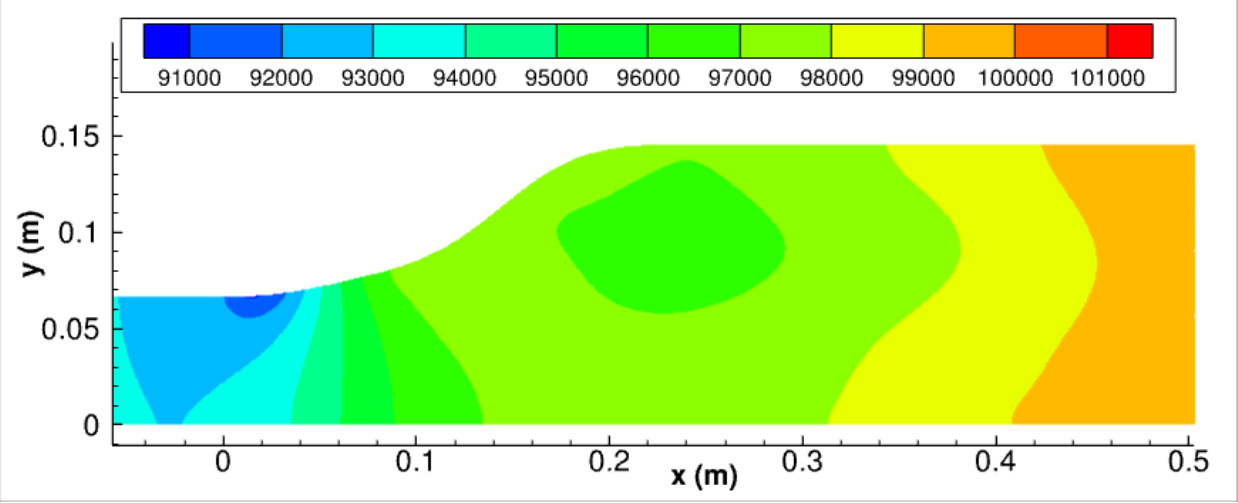}\\
\multicolumn{2}{c}{(a) No control} \\
\includegraphics[clip=true,trim=02 02 02 02, width=0.5\textwidth]{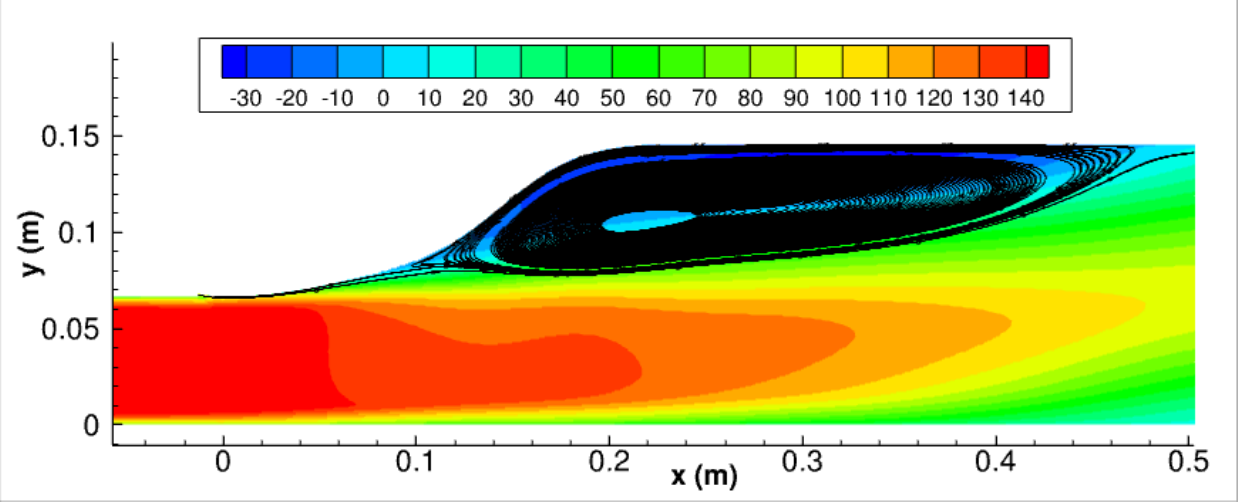}&
\includegraphics[clip=true,trim=02 02 02 02, width=0.5\textwidth]{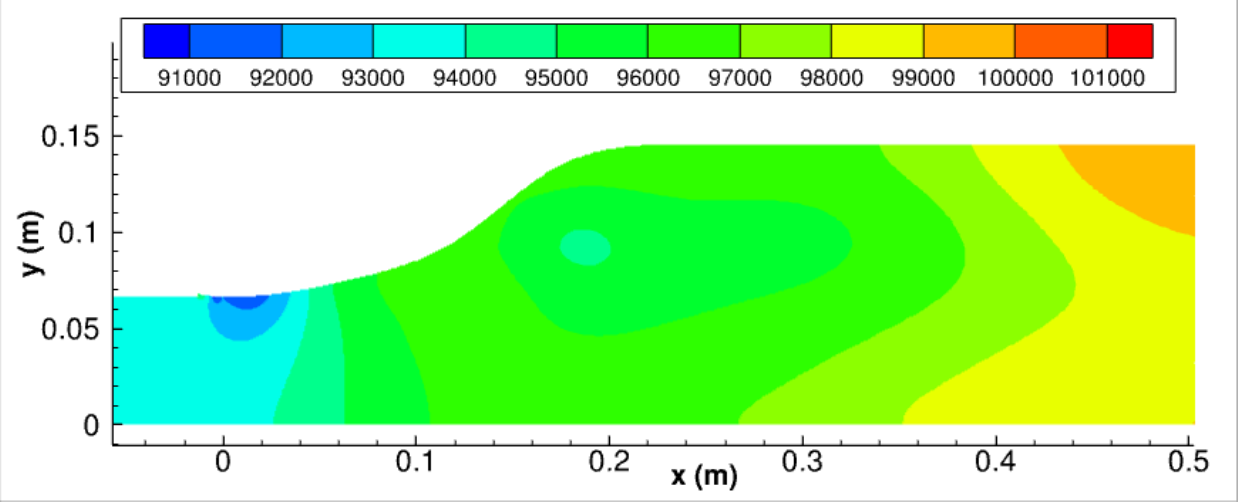}\\
\multicolumn{2}{c}{ (b) Steady jet control}\\
  \includegraphics[clip=true,trim=02 02 02 02, width=0.5\textwidth]{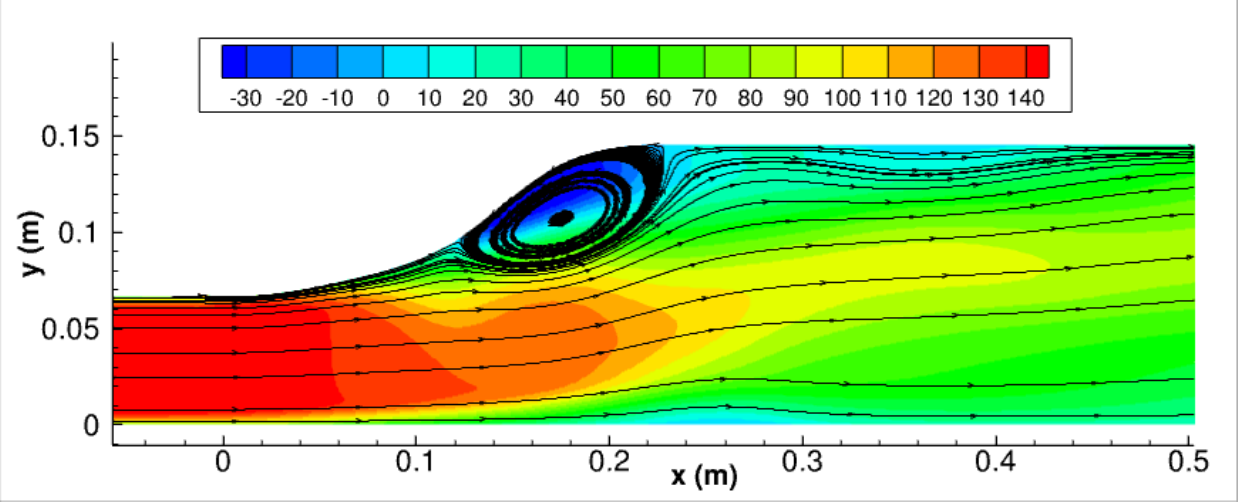}&
 \includegraphics[clip=true,trim=02 02 02 02, width=0.5\textwidth]{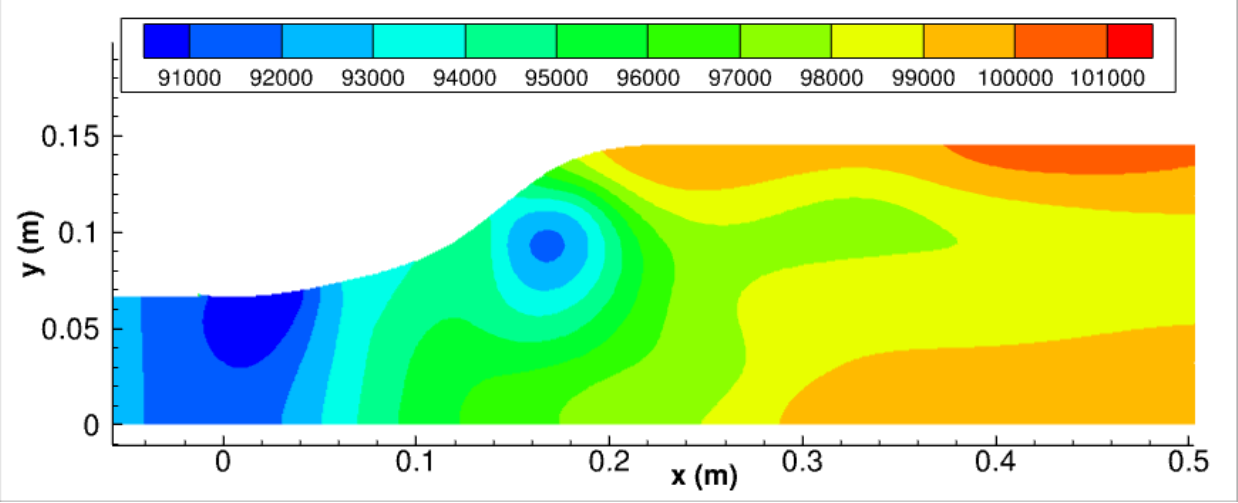}\\
\multicolumn{2}{c}{ (c) Unsteady high-amplitude jet control with 200 Hz}\\
  \includegraphics[clip=true,trim=02 02 02 02, width=0.5\textwidth]{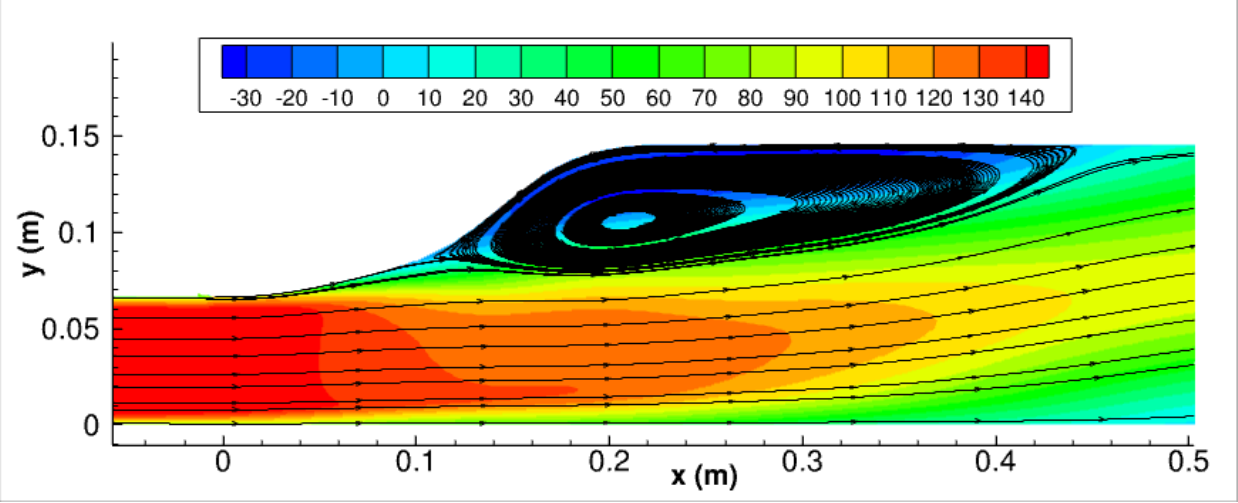}&
  \includegraphics[clip=true,trim=02 02 02 02, width=0.5\textwidth]{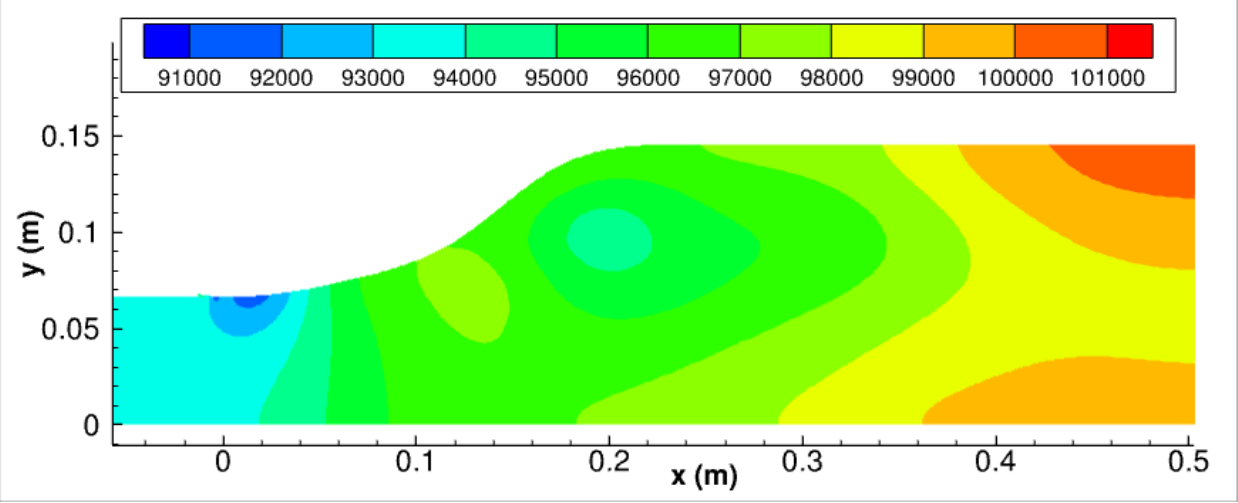}\\
\multicolumn{2}{c}{ (d) Unsteady low-amplitude jet control with 200 Hz}\\
\end{tabular}
\caption{Spatial distribution of the velocity (m/s) (left), along with black streamlines, and the pressure field (Pa) (right) near the Aerodynamic Interface Plane (AIP) using the 2-D model at 1 s.}
\label{fig:2DSpatialPVel}
\end{figure}

Figure~\ref{fig:2DSpatialPVel} shows the spatial distribution of the velocity field and pressure field near the recirculation region for the baseline (no control) and controlled cases with steady and pulsed jets with two different RMS at 200 Hz. As shown, the adverse pressure gradient due to the deflection of the flow near the diffuser inlet results in boundary layer separation, which, in turn, results in less pressure recovery. Using a steady jet with a mass flow rate of 1\% of the diffuser mass flow rate results in larger pressure values at the ($x=0.25$~m) constant plane, so larger pressure recovery. The size of the separation decreases slightly. However, as shown in Fig.~\ref{fig:2DSpatialPVel}(c), the recirculation region shrinks significantly with high RMS, and pressure recovery near the wall increases. The lower amplitude jet does not change the size of the recirculation region, as can be seen by comparing Fig.~\ref{fig:2DSpatialPVel}(a) with (d). This important finding shows that adaptive and active flow control is very sensitive, especially for highly efficient designs.

\begin{figure}
\centering
\begin{tabular}{c}
\includegraphics[clip=true,trim=02 02 02 02, width=1.05\textwidth]{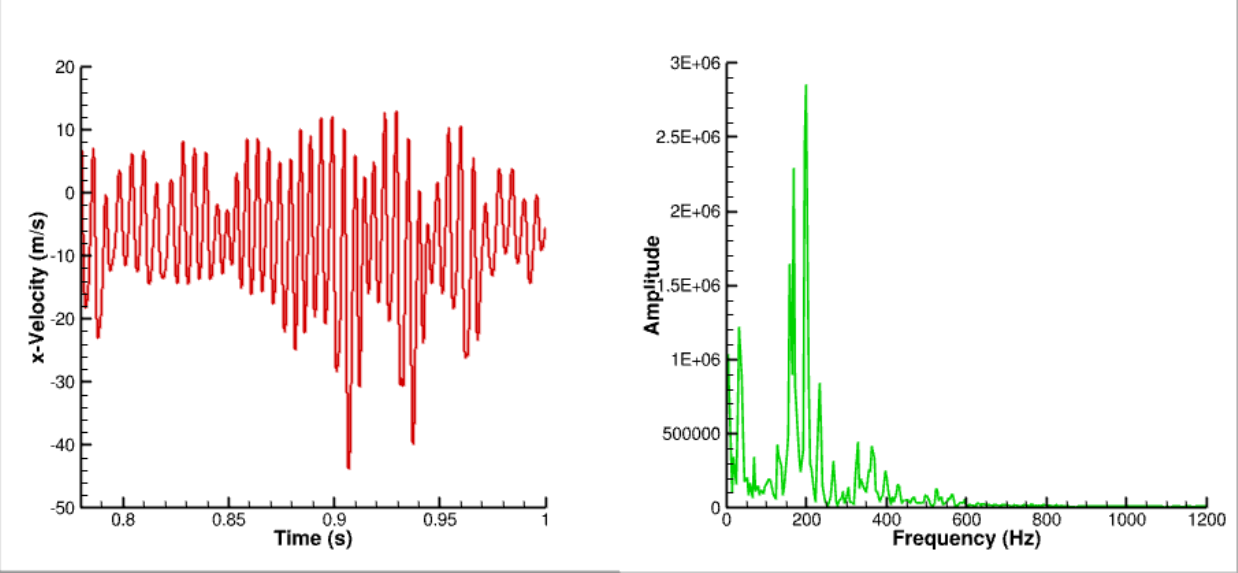}\\
(a) Control with the steady jet \\
\includegraphics[clip=true,trim=02 02 02 02, width=1.05\textwidth]{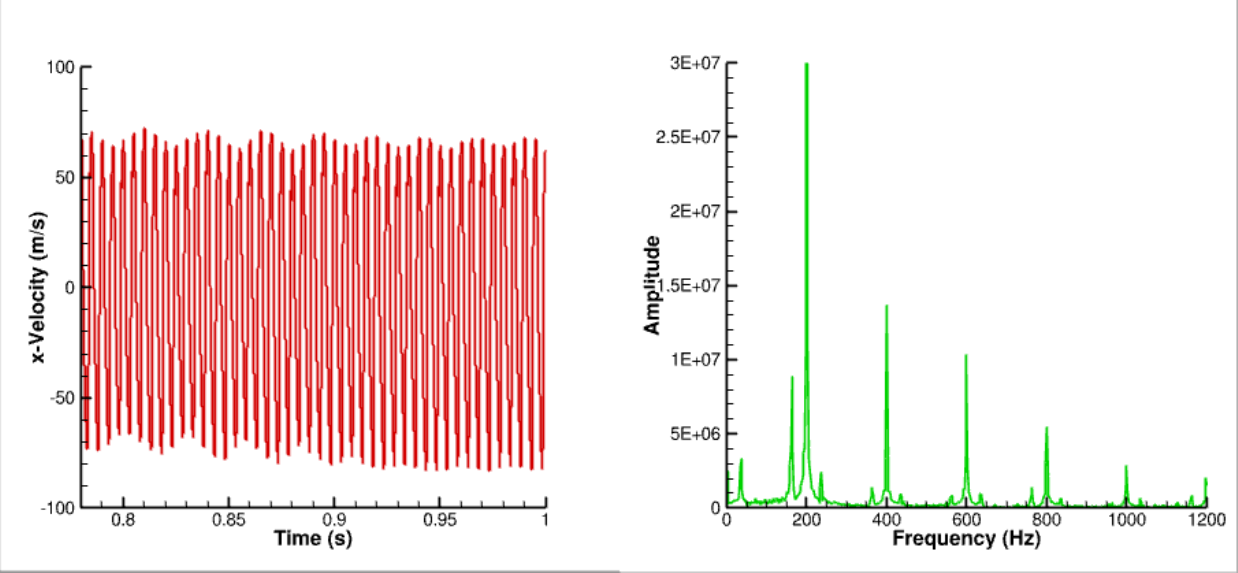}\\
(b) Control with the unsteady high-RMS jet \\
\end{tabular}
\caption{Time history of the axial velocity inside the separation bubble (left) at $(x,y)$ = (0.1761, 0.1084)~m and the corresponding frequency spectrum (right). }
\label{fig:2DFFT}
\end{figure}

Figure~\ref{fig:2DFFT} shows the complex flow nature that introduces various frequencies along with the unsteady jet frequencies. For the steady and unsteady jets, the dominant frequency is found to be 200 Hz inside the separation region. Therefore, our controller starts with this frequency and finds a better value to increase the pressure recovery.

\begin{figure}
\centering
\begin{tabular}{cc}
\includegraphics[clip=true,trim=02 02 02 02, width=0.5\textwidth]{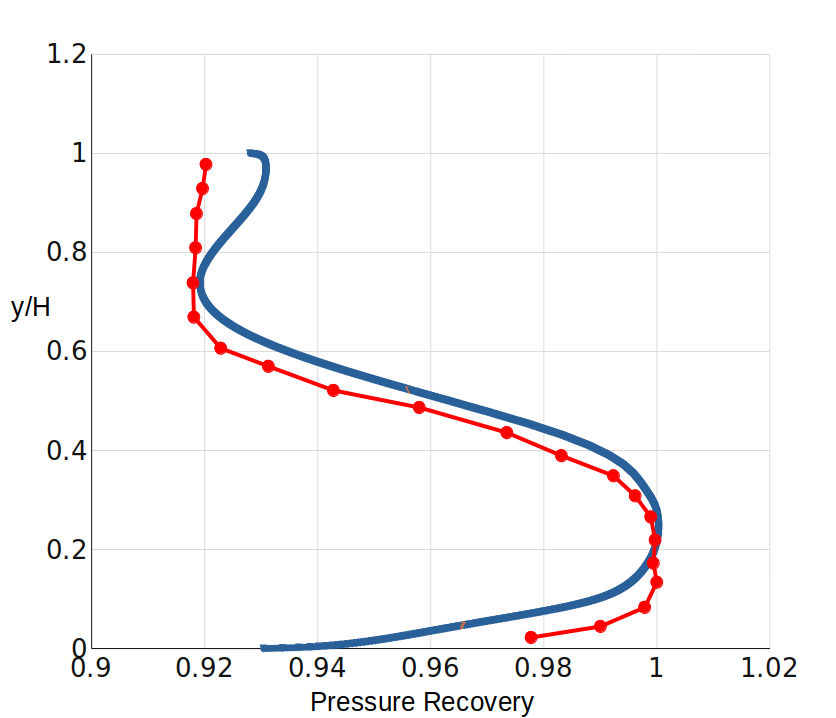}&
\includegraphics[clip=true,trim=02 02 02 02, width=0.5\textwidth]{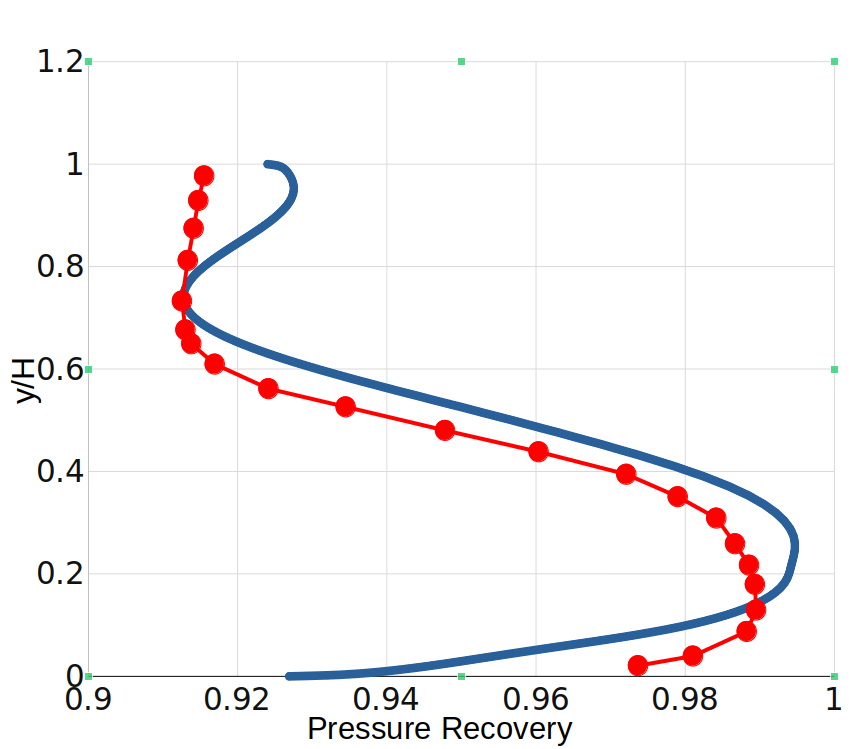}\\
\end{tabular}
\caption{Comparison of the calculated (blue) pressure recoveries with the experiment (red)~\cite{gartner2019mitigation} along the AIP plane for steady (left) and low-RMS jet (right) control case.}
\label{fig:2DComparisonExp}
\end{figure}

To ensure the accuracy of the computations, we validate our results thoroughly with the experiments for complex unsteady internal flows with a Reynolds number of $1.4 \times 10^6$. Figure~\ref{fig:2DComparisonExp} compares the pressure recovery along the Aerodynamic Interface Plane (AIP), located at $x=$0.25m constant plane. Except for the recirculation region, the calculated pressure recovery values for steady and low-RMS cases are in good agreement with the experiment~\cite{gartner2019mitigation}, indicating that the current calculations are performed accurately. The discrepancies near the walls can be attributed to experimental uncertanity caused by reflections from the walls. Nonetheless, as will be discussed next, the active controller reduces the size of the recirculation region and prevents 3-D secondary structures in higher–pressure-recovery cases. Therefore, 2-D modeling does not introduce significant errors.

\section{Data-driven Control for Pressure Recovery}
\label{sec:RCAC_Algo}

This section briefly describes the data-driven, learning-based adaptive control framework based on retrospective cost optimization to maximize the pressure recovery in the supersonic diffuser by modulating the jet velocity frequency. 
Since a single performance metric (pressure recovery) is considered and only one signal (jet velocity frequency) is modified, the system, from a control perspective, is single-input, single-output (SISO).
While the general control framework is provided in this section, specific details for implementation with SISO systems are also given when relevant.
%
%
RCAC is described in detail in \cite{rahmanCSM2017} and its extension to digital PID control is given in \cite{rezaPID}.
While RCAC is used for online optimization in \cite{paredes2025combustor}, the RCAC algorithm used in this work is based on the algorithms shown in \cite{paredes2023rijke}, \cite{paredes2025suppression}.
Subsection \ref{subsec:rcac} provides a review of RCAC. 
Subsection \ref{subsec:ctrl_impl} introduces the sampled-data implementation of RCAC to interface the discrete-time controller with a continuous-time system.
Subsection \ref{subsec:implementation_details} provides the implementation details for interfacing RCAC with the diffuser model simulation.

\subsection{Review of Retrospective Cost Adaptive Control} \label{subsec:rcac}

Consider a system
\begin{align}
    x_{k+1} &= f_k(x_k,u_k,w_k), 
    \label{eq:state}\\
    y_k     &= g_k(x_k,w_k), 
    \label{eq:output}
\end{align}
where
$x_k \in \BBR^{l_x}$ is the state,
$u_k \in \BBR^{l_u}$ is the input,
$y_k \in \BBR^{l_y}$ is the measured output, and
$w_k \in \BBR^{l_w}$ is the exogenous signal that can represent commands, external disturbance, or both.
The functions $f_k$ and $g_k$ represent the dynamics map and the output map. 
The goal is to develop an adaptive control law that drives the output $y_k$ to a desired values with limited modeling information about \eqref{eq:state}, \eqref{eq:output}.
Note that explicit knowledge of $f_k$ and $g_k$ is not assumed since RCAC requires only the input and output measurements.

Consider the strictly proper, discrete-time, input-output controller  
\begin{align}
    u_{ k} = \sum_{i=1}^{l_\rmc}P_{i,k}u_{k-i} + \sum_{i=1}^{l_\rmc}Q_{i,k}z_{k-i}, \label{IO_controller}
\end{align}
where $u_{ k} \in \mathbb R^{l_u}$
is the control input,  $z_k \in \mathbb R^{l_z}$ is the measured performance variable, $l_\rmc$ is the controller-window length, and, for all $i\in \{1,\ldots,l_\rmc\},$  $P_{i,k} \in \mathbb R^{l_u \times l_u}$ and $Q_{i,k} \in \mathbb R^{l_u \times l_z}$ are the controller coefficient matrices.
%
%
The controller \eqref{IO_controller} can be written as
\begin{align}
    u_{ k}   =   \phi_k  \theta_k , \label{controller}
\end{align}
where 
\begin{align}
	\phi_k &\isdef
    \left[ \arraycolsep=3pt\def\arraystretch{0.9} \begin{array}{cccccc} 
    			u_{k-1}^\rmT & \cdots & u_{k-l_\rmc}^\rmT & z_{k-1}^\rmT & \cdots & z_{k-l_\rmc}^\rmT
    \end{array} \right]
    		\otimes
    		I_{l_u}
    		\in \mathbb{R}^{l_u \times l_{\theta}},  \label{controller_phi} \\
    	\theta_k &\isdef {\rm vec}
    \left[ \arraycolsep=1.7pt\def\arraystretch{0.9} \begin{array}{cccccc} 
        P_{1,k} &\cdots &P_{l_\rmc,k} &Q_{1,k} &\cdots &Q_{l_\rmc,k}
    \end{array} \right] \in \BBR^{l_\theta},
\end{align}
$l_\theta \isdef l_\rmc l_u (l_u + l_z),$ and $\theta_k$ is the vector of controller coefficients, which are updated at each time step $k$.
Note that, if $z_k$ and $u_k$ are scalar, then the SISO transfer function of  \eqref{IO_controller} from $z_k$ to $u_k$ is given by
\begin{align}
G_{\rmc,k}(\textbf{q})= \frac{Q_{1,k} \textbf{q}^{l_\rmc - 1} + \cdots + Q_{l_\rmc,k} }{\textbf{q}^{l_\rmc} - P_{1,k}\textbf{q}^{l_\rmc - 1} - \cdots - P_{l_\rmc,k} },
\end{align}
where $\textbf{q}\in \BBC$ is the forward-shift operator.

Next, define the retrospective cost variable
\begin{align}
	\hat z_k (\hat \theta) \isdef z_k  - G_\rmf(\textbf{q})(u_k - \phi_k \hat{\theta}), \label{zhat1}
\end{align}
where 
$G_{\rmf}(\shiftq)$ is an $l_z \times l_u$ asymptotically stable, strictly proper transfer function, and
$\hat{\theta} \in \mathbb R^{l_\theta}$ is the controller coefficient vector determined by  optimization below.
The rationale underlying \eqref{zhat1} is to replace the applied past control inputs with the re-optimized control input $\phi_k \hat{\theta}$ 
so that the 
closed-loop transfer function from $u_k - \phi_k \theta_{k+1}$ to $z_k$
matches $G_{\rmf}$ \cite{rahmanCSM2017}.
Consequently, $G_{\rmf}$ serves as a closed-loop target model for adaptation.
In the present paper, $G_{\rmf}$ is chosen to be a finite-impulse-response transfer function of window length $n_\rmf$ of the form
\begin{align}
	G_{\rmf}(\textbf{q}) \isdef
			\sum_{i=1}^{n_\rmf} N_i \textbf{q}^{-i},
\end{align}
where  $N_1,\ldots,N_{n_\rmf}\in\BBR^{l_z\times l_u}.$ 
We can thus rewrite \eqref{zhat1} as 
\begin{align}
    \hat z_k(\hat \theta) = z_k - N ( \bar{U}_k - \Phi_k\hat{\theta} ),
\end{align}
where
\begin{align} \Phi_k \isdef 
\left[ \arraycolsep=0.8pt\def\arraystretch{0.9} \begin{array}{c} 
        \phi_{k-1}       \\
         \vdots          \\
         \phi_{k-n_\rmf} \\
    \end{array}  \right] \in \BBR^{ n_\rmf l_u  \times l_\theta },\
    U_k \isdef 
\left[ \arraycolsep=0.8pt\def\arraystretch{0.9} \begin{array}{c} 
        u_{k-1}       \\
        \vdots        \\
        u_{k-n_\rmf}  \\
    \end{array}   \right] \in \BBR^{ n_\rmf l_u   },
\end{align}
\begin{align}
    N \isdef  \left[\  N_1  \ \cdots  \  N_{n_\rmf} \   \right] \in \BBR^{l_z \times n_\rmf l_u }.
\end{align}

The choice of $N$ includes all required modeling information.
When the plant is SISO, that is, $l_z=l_u=1,$ this information consists of the sign of the leading numerator coefficient, the relative degree of the sampled-data system, and all nonminimum-phase (NMP) zeros \cite{rahmanCSM2017}.
Since zeros are invariant under feedback, omission of a NMP zero from $G_\rmf$ may entail unstable pole-zero cancellation.
Cancellation can be prevented, however, by using the control weighting $R_u$ introduced below, as discussed in \cite{rahmanCSM2017}.
%
%

Using \eqref{zhat1}, we define the  cumulative cost function
\begin{align}
    J_k(\hat{\theta}) &\isdef \sum\limits_{i=0}^{k} [  \hat z_i^{\rm T}(\hat \theta) \hat z_i(\hat \theta)  +   (\phi_i \hat \theta)^{\rm T}  R_u \phi_i \hat \theta ] 
    +   (\hat \theta - \theta_0 ) ^{\rm T}   P_0^{-1} (\hat \theta - \theta_0 ), \label{eq:Jg}
\end{align}
%
where $P_0 \in \BBR^{l_\theta \times l_\theta}$ is positive definite, and $R_u \in \BBR^{l_u \times l_u}$ is positive semidefinite. 
As can be seen from \eqref{eq:Jg}, $R_u$ serves as a control weighting, and the matrix $P_0^{-1}$ defines the regularization term and initializes the recursion for $P_k$ defined below.
The following result uses recursive least squares (RLS) 
\cite{ljung1983,islam2019recursive} to minimize \eqref{eq:Jg}, where, at each step $k,$  the minimizer of \eqref{eq:Jg} is the update $\theta_{k+1}$ of the controller coefficient vector.


\begin{proposition}
For all $k \ge 0$,
let $\theta_{k+1}$ denote the global minimzer of $J_k(\hat \theta)$ given by \eqref{eq:Jg}.
Then, $\theta_{k+1}$ is given by
\begin{align}
    \theta_{k+1} &= \theta_k  - P_{k+1}             
\left[ \arraycolsep=0.9pt\def\arraystretch{0.9} \begin{array}{c} 
                N\Phi_k \\
                \phi_k 
            \end{array}\right]^\rmT \hspace{-0.5em} \bar{R}             
\left[ \arraycolsep=0.9pt\def\arraystretch{0.9} \begin{array}{cc} 
                 z_k - N ( U_k - \Phi_k \theta_k ) \\
                \phi_k\theta_k
            \end{array}\right] \label{eq:theta_update},
\end{align}
where
%
%
%
\begin{align}
   \Gamma_k &\isdef
                 \bar{R} - 
                    \bar{R}
                    \left[ \arraycolsep=1.1pt\def\arraystretch{0.9} \begin{array}{c} 
                N\Phi_k \\
                \phi_k 
            \end{array}\right]
            \left(P_k^{-1} +
            \left[ \arraycolsep=1.1pt\def\arraystretch{0.9} \begin{array}{c} 
                N\Phi_k \\
                \phi_k 
            \end{array}\right]^\rmT 
            \bar{R}
            \left[ \arraycolsep=1.1pt\def\arraystretch{0.9} \begin{array}{c} 
                N\Phi_k \\
                \phi_k 
            \end{array}\right]
            \right)^{-1}
            \left[ \arraycolsep=1.1pt\def\arraystretch{0.9} \begin{array}{c} 
                N\Phi_k \\
                \phi_k 
            \end{array}\right]^\rmT 
            \bar{R} 
            \in \BBR^{ (l_z + l_u) \times  (l_z + l_u) } ,\\
    \bar{R} &\isdef {\rm diag}( I_{l_z} , R_u ) \in \BBR^{ (l_z + l_u) \times  (l_z + l_u) },
\end{align}
and $P_k$ satisfies
\begin{align}
    P_{k+1}      &=  P_k  - P_k 
    \left[ \arraycolsep=0.9pt\def\arraystretch{0.9} \begin{array}{c} 
                N\Phi_k \\
                \phi_k 
            \end{array}\right]^\rmT
    \hspace{-0.5em} \Gamma_k
    \left[ \arraycolsep=0.9pt\def\arraystretch{0.9} \begin{array}{c} 
                N\Phi_k \\
                \phi_k 
            \end{array}\right]
    P_k , \label{eq:pk_update}
\end{align}
\end{proposition}
\textbf{Proof:} See \cite{goel2020recursive}. \hfill $\square$

Hence, RCAC is updated by \eqref{eq:pk_update}, \eqref{eq:theta_update}, and the control input is calculated with \eqref{controller}.
For all of the numerical simulations in this paper, $\theta_k$ is initialized as $\theta_0=0_{l_\theta\times 1}$ to reflect the absence of a prior baseline controller.
Furthermore, $G_\rmf(\textbf{q}) \in \{ -1/{\textbf{q}}, \ 1/{\textbf{q}} \}$,
that is, $N$ is either $+1$ or $-1$.
The sign of $N$, in this case, reflects an initial search direction in the $u_k$ space.
Aside from the hyperparameter selection discussed above, RCAC uses no other modeling information.
For convenience, we set  $P_0 = p_0 I_{l_\theta},$ where the scalar $p_0>0$ determines the initial rate of adaptation.

\subsection{Sampled-data Controller Implementation} \label{subsec:ctrl_impl}

\begin{figure} [h!]
    \centering
    \resizebox{0.6\columnwidth}{!}{%
    \begin{tikzpicture}[>={stealth'}, line width = 0.25mm]

    \node [input, name=ref]{};
    \node [sum, right = 0.75cm of ref, inner sep = 0.001em] (sum2) {};
    \node[draw = none] at (sum2.center) {$+$};
    \node [smallblock, rounded corners, right = 0.75cm of sum2 , minimum height = 0.6cm , minimum width = 0.7cm] (error_norm) {$\SN$};
    %
    %
    \node [smallblock, rounded corners, right = 0.75cm of error_norm , minimum height = 1cm , minimum width = 0.7cm] (controller) {$G_{\rmc,k}$};
    %
    %
    \node [smallblock, rounded corners, right = 0.75cm of controller, minimum height = 0.6cm , minimum width = 0.5cm] (DA) {\scriptsize ZOH};
    \node [smallblock, fill=green!20, rounded corners, right = 0.75cm of DA, minimum height = 0.6cm , minimum width = 1cm] (system) {$\SM$};
    \node [output, right = 0.5cm of system] (output) {};
    \node [input, below = 1.25cm of system] (midpoint) {};
    
    \draw [->] (controller) -- node [above] {$u_{ k}$} (DA);
    \draw [->] (DA) -- node [above] (du) {$u$} (system);

    \node[circle,draw=black, fill=white, inner sep=0pt,minimum size=3pt] (rc11) at ([xshift=-2.9cm]midpoint) {};
    \node[circle,draw=black, fill=white, inner sep=0pt,minimum size=3pt] (rc21) at ([xshift=-3.2cm]midpoint) {};
    \draw [-] (rc21.north east) --node[below,yshift=.55cm]{$T_\rms$} ([xshift=.3cm,yshift=.15cm]rc21.north east) {};
    
    \draw [->] (system) -- node [name=y, near end]{} node [very near end, above] {$y$}(output);
    
    \draw [->] (ref.east) -- node [near start, above,xshift = 0.1cm] {$r_k$}  node [near end, above] {} (sum2);
    \draw [-] (y.west) |- (midpoint);
    \draw [-] (midpoint) -| node [very near end, above, xshift=-0.7cm] {$y_k$} (rc11.east);
    \draw [->] (rc21) -| node [very near end, xshift = -0.35cm, yshift = 0.1cm] {\huge -} (sum2.south);
    \draw [->] (sum2.east) -- node [above, xshift = -0.05cm] {$e_k$} (error_norm.west);
    \draw [->] (error_norm.east) -- node [above, xshift = -0.05cm] {$z_k$} (controller.west);
    
    \end{tikzpicture}
    }  
    \caption{Sampled-data implementation of adaptive controller for control of the continuous-time system $\SM.$ 
    For this work, $\SN$ is the normalization function \eqref{error_norm}, 
    and $\SM$ represents the diffuser model described in Section \ref{sec:model}.
    }
    \label{AC_CT_blk_diag}
\end{figure}
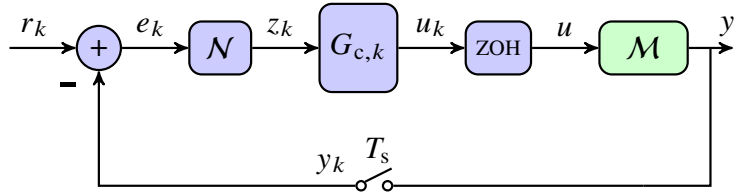

The adaptive controller is implemented as a sampled-data controller. 
Figure \ref{AC_CT_blk_diag} shows a block diagram of the sampled-data closed-loop system, where $y \in \BBR$ is the output of the continuous-time system $\SM$, $y_k$ is the sampled output, $r_k \in \BBR$ is the discrete-time command,  $e_k \isdef r_k - y_k$ is the command-following error, and $T_\rms>0$ is the sampling period.
The digital-to-analog (D/A) and analog-to-digital (A/D) interfaces, which are synchronous, are zero-order-hold (ZOH) and sampler, respectively.
Finally, $\SM$ represents the diffuser model introduced in Section \ref{sec:model}.

In this work, the performance variable $z_k$, which is used for adaptation, is the normalized error  
\begin{align}
    \SN(e_k) \isdef \frac{K_\rme e_k}{1 + \nu |K_\rme e_k|}, \label{error_norm}
\end{align}
where $K_\rme$ is an error gain and $\nu\in[0,\infty).$ 
We fix $\nu = 0.2$ throughout the paper.
The adaptive controller $G_{\rmc, k}$ operates on $z_k$ to produce the discrete-time control $u_{ k} \in \BBR.$ 
%
%
%
%
Implementation of the adaptive controller requires selection of the hyperparameters $l_\rmc$, $p_0,$ $R_u,$ $N,$ and $K_\rme$ depending on the system and performance requirements, which is described next.
%
%

\subsection{Implemetation Details for Interface with Diffuser Model Simulation} \label{subsec:implementation_details}

For the RCAC measurement $y,$ we consider the pressure recovery array $p_{\rmr} \in \BBR^{l_p}$ whose elements correspond to pressure recovery values sampled within a plane.
In particular, a custom pressure-recovery sampling utility is developed within the OpenFOAM framework.
The utility reads the time-averaged pressure, temperature, and velocity fields at user-specified locations and first computes the inlet total pressure, which is taken as the reference stagnation pressure. 
The same procedure is then applied to the Aerodynamic Interface Plane (AIP), where the local stagnation pressure and the corresponding pressure-recovery field are evaluated, which are provided in $p_{\rmr}.$
%
%
%
Next, let $p_{\rmr, {\rm avg}} \in \BBR$ be the average of all the components in $p_{\rmr}.$
Hence, RCAC uses $p_{\rmr, {\rm avg}}$ as a measurement, such that $y = p_{\rmr, {\rm avg}}.$
Furthermore, for this work, we set $r_k \equiv 1$ to maximize pressure recovery within the diffuser model.

In this work, RCAC modulates the jet frequency.
Thus, 
\begin{equation}
    \mathbf{V}(t) = \mathbf{L} + A \cdot \sin(2\pi \  f(t) \ t) \cdot \mathbf{S},
    \label{Eq:Jetparam_mod}
\end{equation}
where, for $t \in [kT_\rms, (k+1)T_\rms),$
\begin{align}
    f(t) = f_\rmb + u_k,
\end{align}
and $f_\rmb = 200 $ $\rm Hz$ is a constant baseline frequency.
%
The values for the parameters $f_\rmb, \mathbf{L}, \mathbf{S}, A$ correspond to the High-Amplitude case shown in Table \ref{tab:jetParameters}, with $f_\rmb = 200$ Hz.
%

The operation of RCAC and the chosen hyperparameters are discussed next. 
RCAC updates the frequency at intervals of $T_\rms = 0.05$ s.
In this work, RCAC is enabled at $t = 0.2$ s after the simulation starts, such that $u_k = 0$ for all $t < 0.2$ s. 
Then, for all $t \ge 0.2$ s, RCAC obtains the $p_{\rmr, {\rm avg}}$ measurement and updates jet frequency $f(t).$
%
%
%
The RCAC hyperparameters are 
\begin{equation*}
    l_\rmc = 5, \ p_0 = 100, \ R_u = 0.25, \ K_\rme = 10.
\end{equation*}
Note that $R_u$ is scalar since the system is SISO and thus $l_y = l_u = 1.$
In the case of $N,$ both values are used in different simulation cases, as discussed below.

In this work, three cases are run. 
A case without RCAC, that is, $u_k = 0$ for all $k \ge 0$, is run to assess the performance improvement from RCAC.
Two cases are run with RCAC as discussed above, with $N = 1$ and $N = -1,$ which allow RCAC to follow two different search directions in the $u_k$ space.

\section{Results from Numerical Simulations with Data-Driven Controller} \label{sec:results}

\begin{figure}[h!]
\centering
\includegraphics[width=0.6\columnwidth]{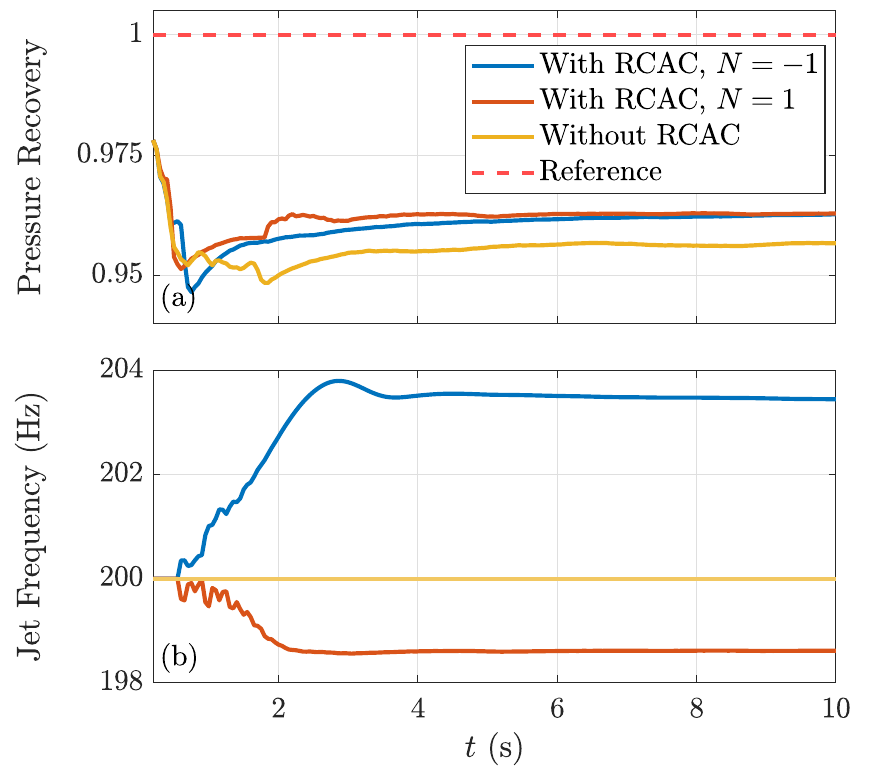}
\caption{
Results of the numerical simulations with and without RCAC. 
(a) shows the measured average pressure recovery $p_{\rmr, {\rm avg}}$ and (b) shows the jet frequency $f$ versus time for the case without RCAC and the cases with RCAC and $N \in \{-1, 1\}.$
}
\label{fig:ex_OpenFOAM_rcac_PR_comparison}
\end{figure}

\begin{figure}[h!]
\centering
\includegraphics[width=0.85\columnwidth]{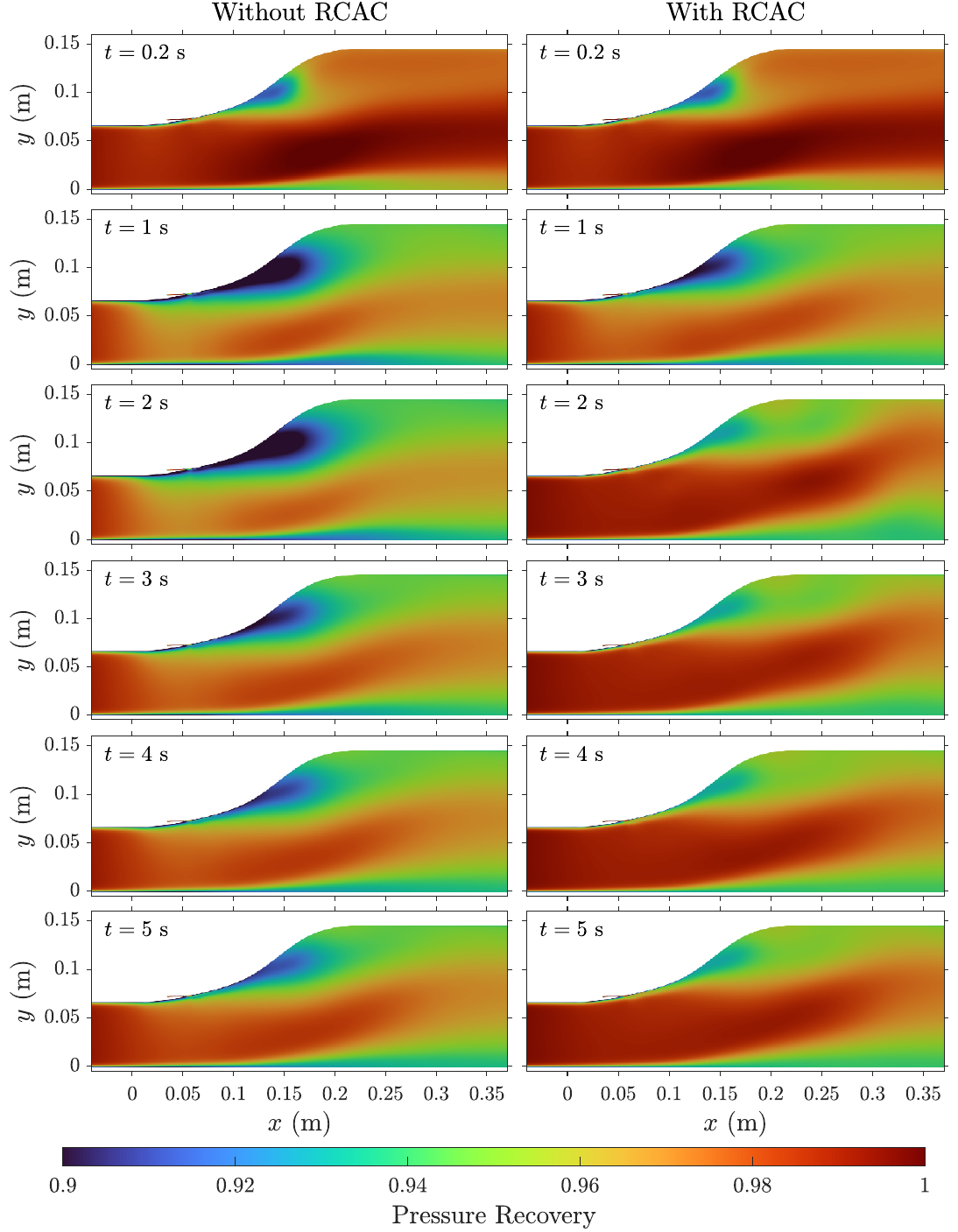}
\caption{
Pressure recovery results from numerical simulations for the baseline case without RCAC and the RCAC-enabled case with $N= 1.$
}
\label{fig:ex_OpenFOAM_rcac_PR}
\end{figure}

\begin{figure}[h!]
\centering
\includegraphics[width=0.85\columnwidth]{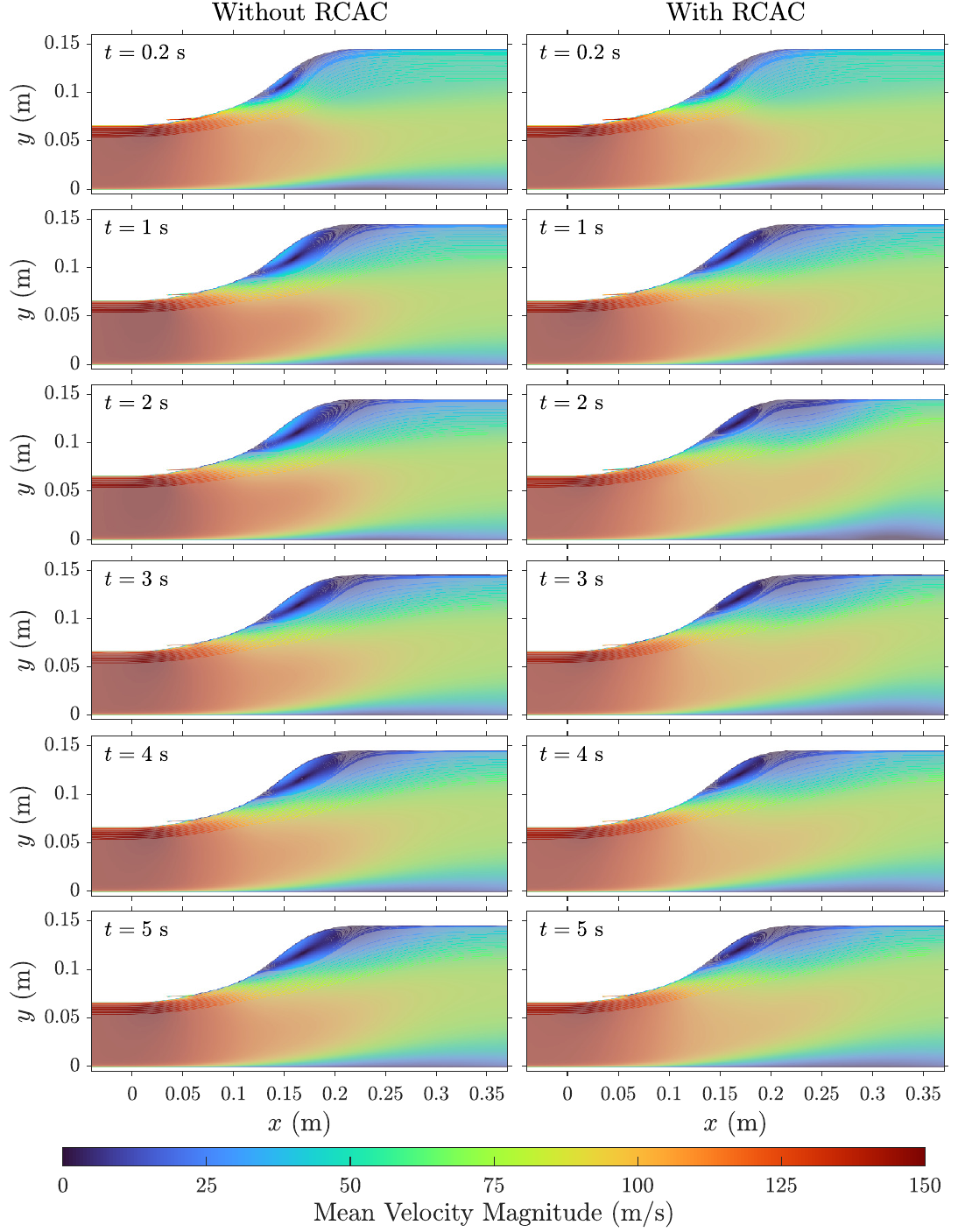}
\caption{
Mean velocity magnitude from numerical simulations for the case without RCAC and the case with RCAC using $N= 1.$
The streamlines corresponding to the air flow over the upper surface of the diffuser are also shown.
}
\label{fig:ex_OpenFOAM_rcac_Umean}
\end{figure}

The numerical simulation results with and without RCAC are shown in Figures \ref{fig:ex_OpenFOAM_rcac_PR_comparison}, \ref{fig:ex_OpenFOAM_rcac_PR}, and \ref{fig:ex_OpenFOAM_rcac_Umean}.
%

%
Figure \ref{fig:ex_OpenFOAM_rcac_PR_comparison} shows the evolution of average pressure recovery $p_{\rmr, {\rm avg}}$ and the jet frequency $f$ in the baseline case without RCAC and in the cases with RCAC and $N \in \{-1, 1\},$ which illustrate the performance from the perspective of the controller, since it interacts with the system solely through these variables.
These plots show that, while both RCAC-enabled cases converge to difference jet injection frequencies, these achieve higher measured avergae pressure recovery than the baseline case without RCAC.

%
%
Figure \ref{fig:ex_OpenFOAM_rcac_PR} shows snapshots of the instantaneous pressure recovery distribution within the diffuser for the baseline case without RCAC and the RCAC-enabled case with $N = 1.$
In the baseline case, the diffuser experiences a persistent low pressure region near the upper wall, as indicated by the extended blue and green regions corresponding to values of pressure recovery lower than 0.95.
In contrast, the RCAC-enabled case shows an improvement in the pressure recovery distribution; while a small low pressure region forms initially, its extent is substantially reduced and does not intensify in time as observed in the baseline case.
Starting from $t = 1$ s, the pressure field becomes dominated by yellow and red regions, and, by $t \in [3, 5]$ s, the pressure field is more dominated by red regions closer to a pressure recovery of 1 than the baseline case, indicating that the implementation of RCAC for frequency modulation effectively increases pressure recovery throughout the diffuser.

%
%
To further examine the effects of the implementation of RCAC, Figure \ref{fig:ex_OpenFOAM_rcac_Umean} shows snapshots of the instantaneous mean velocity magnitude distribution within the diffuser for the same case and time instants as shown in Figure \ref{fig:ex_OpenFOAM_rcac_PR}.
Furthermore, the streamlines corresponding to the air flow from the upper surface of the diffuser are also shown.
Consistent with the results shown in Figure \ref{fig:ex_OpenFOAM_rcac_PR}, the baseline case exhibits a large recirculation region that originates within the diffuser expansion and persists throughout the simulation.
This results in a flow separation region, characterized by low local streamwise velocity and prominent backflow, as observed in the strealines.

In contrast, the RCAC-enabled case shows a noticeable reduction in the size of the recirculation region in the upper portion of the diffuser. 
Similarly to the results shown in Figure \ref{fig:ex_OpenFOAM_rcac_PR}, while some degree of flow separation forms initially, the extent of the region is substantially reduced once RCAC starts adapting.
In particular, by $t \in [3, 5]$ s, the streamlines attach further donwstream and the reverse-flow pocket becomes visibly smaller relative to the baseline case, indicating that the implementation of RCAC for frequency modulation effectively reduces flow separation.

These results show that RCAC can improve flow attachment and minimize pressure losses in diffusers.
We emphasize that RCAC achieves the reduction in flow separation without detailed modeling information.
This highlights the potential of adaptive data-driven control to achieve targeted aerodynamic benefits without relying on detailed flow-model knowledge.

\section{Conclusions} \label{sec:conclusions}

This work demonstrated the application of data-driven, adaptive control to improve pressure recovery in an S-shaped diffuser via high-fidelity CFD simulations.
First, steady and unsteady jet-based flow-control strategies were evaluated to characterize the sensitivity of diffuser performance to actuation parameters. 
These preliminary studies highlighted the inherent unsteadiness and strong recirculation present in the diffuser, as well as the potential for unsteady high-amplitude forcing to substantially reduce separation and improve pressure recovery.

A data-driven controller, based on the retrospective cost adaptive control (RCAC) was then integrated with the numerical solver to optimize the jet frequency in real time.
The data-driven controller required minimal modeling information and operated solely on pressure-recovery measurements extracted from the simulation. 
Results show that RCAC consistently improved diffuser performance: both RCAC-enabled cases achieved higher pressure recovery than the baseline configuration.
Furthermore, they converged to distinct injection frequencies, demonstrating the controller's ability to discover effective jet injection frequencies autonomously.
Spatial fields of pressure recovery and mean velocity further revealed that RCAC reduced the size of the recirculation zone near the upper diffuser surface, leading to more uniform and efficient flow in the diffuser.

This numerical investigation suggests that data-driven adaptive control is a promising tool for active flow control in complex internal flows. 
Future work will extend the numerical experiments described in this paper to full three-dimensional simulations, where spanwise instabilities are expected to play a significant role. 
%



\section*{Acknowledgments}
OT is thankful for computational resources from the Center for Computational Innovations at Rensselaer Polytechnic Institute and for computational resources granted by NSF-ACCESS for the project PHY240018.  Financial support for OT for this research was provided by RPI.
AG acknowledges computational resources granted by NSF-ACCESS for the project MCH250107. 

\bibliography{OT_bib}
\end{document}